\documentclass[aip,apl,reprint]{revtex4-1}
\usepackage{graphicx}
\usepackage{dcolumn}
\usepackage{bm}
\usepackage{amsmath}%
\usepackage{amsfonts}%
\usepackage{amssymb}
\begin{document}

\title{Temperature performance analysis of terahertz quantum cascade lasers: Vertical versus diagonal designs }

\author{Alp\'ar M\'aty\'as}
\email{alparmat@mytum.de}
\affiliation{Emmy Noether Research Group ``Modeling of Quantum Cascade Devices``, Technische Universit\"{a}t M\"{u}nchen, D-80333 Munich, Germany}
\affiliation{Institute for Nanoelectronics, Technische Universit\"{a}t M\"{u}nchen, D-80333 Munich, Germany}
\author{Mikhail A. Belkin}
\affiliation{Department of Electrical and Computer Engineering, The University of Texas at Austin, Austin, TX 78758 USA}
\author{Paolo Lugli}
\affiliation{Institute for Nanoelectronics, Technische Universit\"{a}t M\"{u}nchen, D-80333 Munich, Germany}
\author{Christian Jirauschek}
\affiliation{Emmy Noether Research Group ``Modeling of Quantum Cascade Devices``, Technische Universit\"{a}t M\"{u}nchen, D-80333 Munich, Germany}
\affiliation{Institute for Nanoelectronics, Technische Universit\"{a}t M\"{u}nchen, D-80333 Munich, Germany}

\date{\today, published as Appl. Phys. Lett. 96, 201110 (2010)}

\begin{abstract}
Resonant phonon depopulation terahertz quantum cascade lasers based on vertical and diagonal lasing transitions are systematically compared using a well established ensemble Monte-Carlo approach. The analysis shows that for operating temperatures below 200\,K, diagonal designs may offer superior temperature performance at lasing frequencies of about 3.5\,THz and above; however, vertical structures are more advantageous for good temperature performance at lower frequencies. 
\end{abstract}

\maketitle

Holding great promise as coherent sources in the terahertz and infrared regime, quantum cascade lasers (QCLs) have been subject to continuous optimization with respect to temperature performance and output power. Particularly, room temperature operation in the terahertz regime has been a long-standing goal. Progress has been made by introducing resonant phonon depopulation (RPD) designs,~\cite{2003ApPhL..82.1015W,2005OExpr..13.3331W} reducing the number of wells per device period,~\cite{2007ApPhL..90d1112L} and by applying copper metal-metal waveguides.~\cite{2005OExpr..13.3331W} Optimizations with respect to the diagonality of the lasing transition resulted in a 3.9\,THz QCL operating at a record temperature of 186\,K.~\cite{2009ApPhL..94m1105K} However, efforts in improving the temperature performance of a similar vertical 3.2\,THz structure by introducing some degree of diagonality failed.~\cite{BElkinCapasso2009} The goal of this paper is to clarify these apparently contradicting results and to investigate the role of diagonality for the optimization of RPD terahertz QCLs with respect to temperature operation.

We have designed and analyzed three-quantum-well terahertz QCLs with different degrees of diagonality and frequencies between 1.8 and 4.8\,THz. For the self-consistent modeling of QCLs, advanced semiclassical ensemble Monte Carlo (EMC)~\cite{Rossi_New,BonnoDessenne,CallebautHu} or full quantum transport methods~\cite{Rossi_New,AWacker01} are commonly used. Our EMC simulation tool has been specifically developed for the analysis of QCLs,~\cite{2009JAP...105l3102J,2009IJQE...45..1059J} self-consistently including all relevant scattering mechanisms, namely, electron-electron (e-e), electron-longitudinal optical (LO) and acoustic phonon, and electron-impurity scattering. Since interface roughness (IR) highly depends on the growth process and its exact parameters are difficult to measure, this scattering mechanism is included phenomenologically using typical parameter values.~\cite{2009JAP...105l3102J,AWacker01} For the design of GaAs/Al$_{0.15}$Ga$_{0.85}$As THz QCLs, barrier heights between 135-150 meV are commonly used.\cite{2003ApPhL..82.1015W,2005OExpr..13.3331W,2007ApPhL..90d1112L,2009ApPhL..94m1105K,BElkinCapasso2009} For our simulations we assumed a barrier height of 165\,meV, corresponding to 72\% conduction band offset.\cite{bandgp} This somewhat larger value was utilized to avoid sharp anticrossings of the lower laser level with weakly-bound upper states in the second downstream injector, arising in some of the investigated designs and leading to well-known simulation artifacts.\cite{CallebautHu} Effective electron masses of 0.067 in the wells and 0.076 in the barriers are assumed. We verified that the observed trends are robust with respect to the assumed barrier heights and IR values. Our EMC simulation provides self-consistent results for the spectral gain in the terahertz structures, where the gain linewidth is extracted from the above mentioned scattering rates based on lifetime broadening;~\cite{2009JAP...105l3102J} we note that nondiagonal correlation effects can only be considered in fully quantum mechanical approaches.~\cite{AWacker01} 

\begin{table*}
\begin{centering}
\caption{Overview of the designed THz QCLs. All layer thicknesses are given in angstrom, and bold numbers indicate barriers. The underlined wells are doped with a sheet density of 2.7$\times10^{10}$\,cm$^{-2}$ in their 55\,\AA-wide middle region.}
\label{table1}
\begin{tabular*}{\textwidth}{@{\extracolsep{\fill}}ccccc}
\hline \hline 
Freq. & 0\% diagonal & 30\% diagonal & 50\% diagonal & 70\% diagonal\tabularnewline
\hline 
1.8\,THz& \textbf{46}/98/\textbf{31}/76/\textbf{43}/\underbar{161} & \textbf{49}/94/\textbf{35}/78/\textbf{47}/\underbar{161} & \textbf{51}/92/\textbf{39}/80/\textbf{48}/\underbar{161} & \textbf{52}/91/\textbf{48}/80/\textbf{49}/\underbar{161}\tabularnewline
2.3\,THz & \textbf{48}/95/\textbf{27}/73/\textbf{42}/\underbar{157} & \textbf{51}/90/\textbf{31}/77/\textbf{46}/\underbar{158} & \textbf{52}/89/\textbf{35}/80/\textbf{47}/\underbar{159} & \textbf{52}/88/\textbf{42}/81/\textbf{48}/\underbar{160}\tabularnewline
2.8\,THz  & \textbf{48}/94/\textbf{24}/72/\textbf{42}/\underbar{157} & \textbf{51}/89/\textbf{28}/78/\textbf{46}/\underbar{159} & \textbf{52}/87/\textbf{32}/80/\textbf{47}/\underbar{159} & \textbf{53}/86/\textbf{39}/81/\textbf{48}/\underbar{159}\tabularnewline
3.2\,THz  & \textbf{48}/96/\textbf{20}/74/\textbf{42}/\underbar{161} & \textbf{51}/90/\textbf{24}/81/\textbf{46}/\underbar{163} & \textbf{52}/88/\textbf{29}/84/\textbf{47}/\underbar{163} & \textbf{52}/87/\textbf{36}/86/\textbf{48}/\underbar{163}\tabularnewline
4.1\,THz  & \textbf{47}/99/\textbf{15}/73/\textbf{40}/\underbar{164} & \textbf{51}/88/\textbf{19}/83/\textbf{44}/\underbar{164} & \textbf{52}/86/\textbf{23}/87/\textbf{45}/\underbar{164} & \textbf{53}/85/\textbf{32}/89/\textbf{47}/\underbar{164}\tabularnewline
4.8\,THz  & \textbf{49}/98/\textbf{12}/71/\textbf{39}/\underbar{163} & \textbf{52}/86/\textbf{15}/84/\textbf{42}/\underbar{164} & \textbf{53}/83/\textbf{20}/89/\textbf{44}/\underbar{164} & \textbf{54}/82/\textbf{29}/92/\textbf{45}/\underbar{164}\tabularnewline
\hline \hline
\end{tabular*}
\par\end{centering}
\end{table*}

For our analysis, we have designed QCLs with different degrees of diagonality at various operating frequencies. The degree of transition diagonality between upper and lower laser states is quantified in terms of the oscillator strength $f_\mathrm{osc}$, normalized to the value for the corresponding vertical design $f^\mathrm{vert}_\mathrm{osc}$. Thus, for the vertical structures we have $f_\mathrm{osc}/f^\mathrm{vert}_\mathrm{osc}=1$, corresponding to a 0\% diagonal design. A design with $f_\mathrm{osc}/f^\mathrm{vert}_\mathrm{osc}=0.7$ is referred to as a 30\% diagonal structure, etc. In order to make the simulation results comparable, a special effort was made to keep laser designs very similar. In particular, the upper and lower laser level anticrossing energies with injector states (for injection and extraction, respectively) were kept in the range 1.5-1.6\,meV and 3.3-3.6\,meV, respectively. If we assume the Al$_{0.15}$Ga$_{0.85}$As barrier height of 135\,meV, these injection/extraction anticrossing energies become 1.95-2.0\,meV for the injection anticrossing and 4.25-4.6\,meV for the extraction anticrossing, which is in line with the injection/extraction anticrossing values used in the current state-of-the-art devices.\cite{2005OExpr..13.3331W,2007ApPhL..90d1112L,2009ApPhL..94m1105K,BElkinCapasso2009} The energy separations between the upper and lower injector states were kept in the range 37-39\,meV in all laser designs. These parameters combined with the desired laser operating frequency and diagonality of the laser transition uniquely define the layer sequence for the structure.~\cite{2007ApPhL..90d1112L} The layer thicknesses of the designed structures are summarized in Table \ref{table1}. Experimentally, the 0\% diagonal structures were tested for the 2.3\,THz, 3.2\,THz, and 4.1\,THz designs.~\cite{BElkinCapasso2009,Chas10} Also copper double-metal 3.2\,THz QCLs with an active region based on a 30\% diagonal transition were tested experimentally. The devices operated up to 174\,K,~\cite{BElkinCapasso2009} which is lower than the maximum operating temperature of 178\,K for similar devices based on a vertical 3.2\,THz design.~\cite{BElkinCapasso2009,BElkinCapasso2009}

\begin{figure}[pt]
\includegraphics{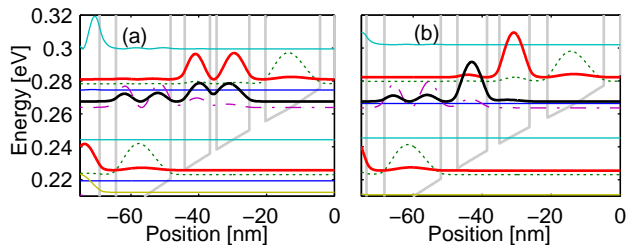}
\caption{(Color online) Conduction band diagram for the 3.2\,THz QCL with (a) 0\% diagonality and (b) 70\% diagonality. Upper and lower laser levels are marked by bold solid lines.}
\label{wfcn}%
\end{figure}

As we go from 0\% diagonal to 70\% diagonal structures, the wavefunctions get more localized in a single well as shown in Fig. \ref{wfcn} for the 3.2\,THz structure. From the basic perspectives of THz QCL design, the diagonal laser transition is expected to help improve the electron injection efficiency into the upper laser state, suppress electron leakage from the upper laser state to the downstream injector, and reduce the nonradiative electron scattering rate from the upper laser state;  however, the diagonal laser transition results in smaller transition dipole moment compared to a vertical transition.~\cite{2009ApPhL..94m1105K,BElkinCapasso2009} One may expect that THz QCL structures based on vertical design may provide higher gain at lower operating temperatures, when electron injection into the upper laser state is efficient and the LO phonon scattering of thermally excited electrons in the upper laser state~\cite{2007NaPho...1..517W} is suppressed. However, diagonal transitions may have advantages at higher operating temperatures. Because the parameters of the electron transport (injection efficiencies and lifetimes in various laser states) are difficult to estimate analytically, we use EMC simulations to compare the peak gain of devices with various diagonalities at different temperatures.

\begin{figure}[pt]
\includegraphics{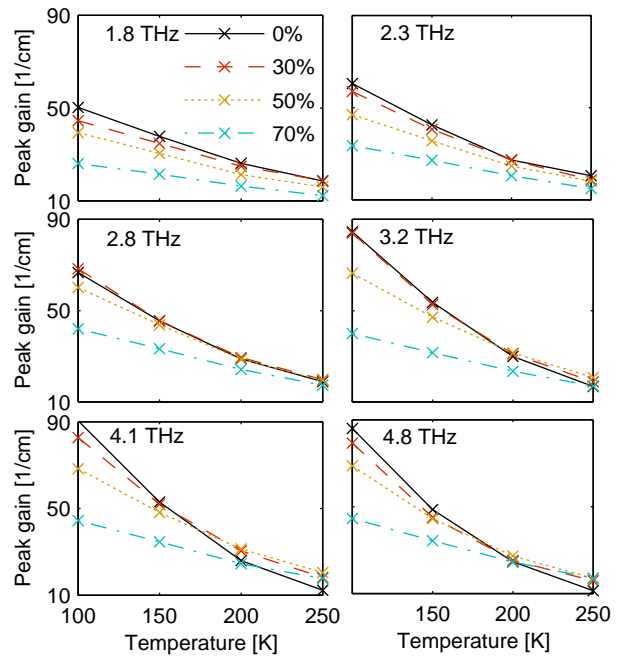}
\caption{(Color online) Dependence of the peak gain on the active region lattice temperature for different operating frequencies and diagonalities. The shown results are for optimum bias, which only weakly depends on temperature and has for each structure been determined at 150\,K as to yield maximum gain.}
\label{diagall}%
\end{figure}

Figure \ref{diagall} shows the temperature dependent peak gain for the designed structures in Table \ref{table1}, as obtained by our EMC simulations. As expected, for increased diagonality, we observe a decreased gain at low temperatures, but also a smaller gain degradation with temperature. At laser lattice temperatures around 150-200\,K, representing the currently relevant range for temperature performance optimizations, diagonality does not offer an advantage for the low frequency structures that we studied. Specifically, in case of the 3.2\,THz design we see no improvement of the temperature performance for the 30\% diagonal as compared to the 0\% diagonal structure, which is consistent with experimental observations.~\cite{BElkinCapasso2009} For higher lasing frequencies, diagonal designs offer clear advantages. The reason is that when the energy spacing between upper and lower laser level increases towards the LO phonon energy (36\,meV in GaAs), scattering of thermally excited electrons in the upper laser level becomes very strong even at modest temperatures,~\cite{2007NaPho...1..517W} resulting in reduced population inversion and thus a strongly decreased peak gain. This detrimental mechanism is suppressed in diagonal designs, which for high operating temperatures maintain increased population difference outweighing their reduced oscillator strength. For low frequency structures, where LO phonon scattering of thermally excited electrons between the laser levels is less pronounced, diagonal designs offer an advantage only at higher operating temperatures, where the thermal activation of LO phonons is stronger. Thus, for the currently reached operating temperatures,~\cite{BElkinCapasso2009,2009ApPhL..94m1105K} we find improvements in peak gain only for designs operating above approximately 3.5\,THz.

\begin{figure}[pt]
\includegraphics{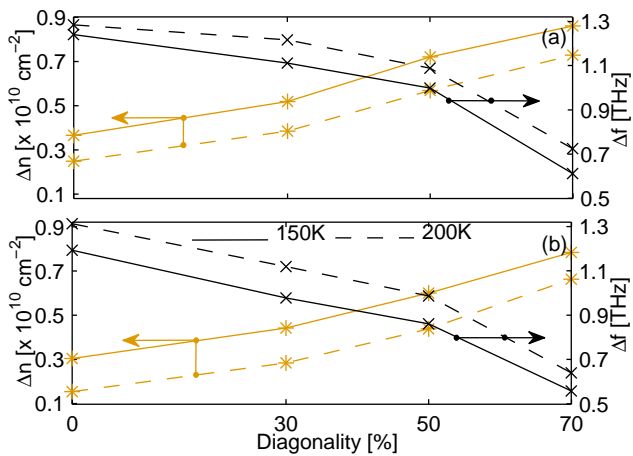}
\caption{(Color online) $\Delta n$ and $\Delta f$ for different degrees of diagonality for the (a) 2.3\,THz QCL and (b) the 4.1\,THz QCL at lattice temperatures of 150 and 200\,K, respectively. The bias is chosen as to maximize the peak gain.}
\label{det}%
\end{figure}

For additional details on our simulation results, we show  in Fig.~\ref{det} the population inversion $\Delta n$ between the upper and lower laser level and gain bandwidth $\Delta f$ (full width at half maximum) at maximum gain, for 2.3\,THz and 4.1\,THz structures of different diagonalities. We note that the peak gain scales with $f_\mathrm{osc}\Delta n/\Delta f$. Overall, the 2.3\,THz structures (Fig.~\ref{det}(a)) exhibit a higher $\Delta f$ in our simulations than the 4.1\,THz designs (Fig.~\ref{det}(b)), which is mainly due to increased Coulomb scattering as well as IR scattering for these structures. Higher $\Delta n$ and, to a lesser extent, smaller $\Delta f$ of more diagonal designs overcompensate the reduction of $f_{osc}$ in case of the 4.1\,THz structures, however not for the 2.3\,THz QCLs. For both structures, a decrease of $\Delta f$ with increasing diagonality is observed. The gain broadening is related to scattering events involving the laser levels. For diagonal lasing transitions, all the contributions get reduced except for IR scattering, which may go up or down. Furthermore, we observe an increase of $\Delta f$ with temperature, which is due to enhanced LO phonon scattering.

\begin{figure}[pt]
\includegraphics{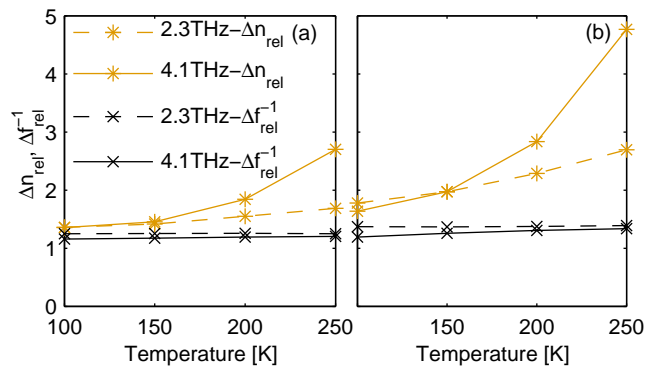}
\caption{(Color online) Temperature dependence for (a) 30\% diagonality and (b) 50\% diagonality of the relative inversion $\Delta n\mathrm{_{rel}}$ and relative inverse linewidth $\Delta f_{\mathrm{rel}}^{-1}$. The peak gain improvement is $\propto \Delta n\mathrm{_{rel}} \Delta f_{\mathrm{rel}}^{-1}$. } 
\label{expl}%
\end{figure}

The effect of diagonality $d$, is further investigated  by introducing the relative quantities $\Delta n\mathrm{_{rel}}=\Delta n(d)/\Delta n(0\%)$ and $\Delta f\mathrm{_{rel}}=\Delta f(d)/\Delta f\mathrm{(0\%)}$. In Fig.~\ref{expl} we plot $\Delta n\mathrm{_{rel}}$ and $\Delta f_{\mathrm{rel}}^{-1}$ as a function of temperature for $d=30\%$ ( Fig.~\ref{expl}(a)) and $d=50\%$ (Fig.~\ref{expl}(b)). Comparing Fig.~\ref{expl}(a) to (b), we find that $\Delta n\mathrm{_{rel}}$ increases with temperature. This trend is most pronounced for 4.1\,THz devices with 50\% diagonality. Very little change of $\Delta f_{\mathrm{rel}}^{-1}$ in the temperature range shown in Fig.~\ref{expl} indicates that diagonal structures at elevated temperatures mainly profit from improved inversion. Finally, Fig.~\ref{expl} clearly shows that diagonal designs offer a much stronger improvement in inversion for the 4.1\,THz QCLs than for the 2.3\,THz structures.

To investigate the robustness of our results, simulations assuming different IR scattering parameters as well as simulations with a fixed lasing transition linewidth (of 1\,THz)  were performed (not shown). They all confirm the previously obtained trend of the spectral peak gains for the different structures. We find that for the case of a fixed linewidth, the gain improvement for diagonal designs is only somewhat reduced.

In summary, we have designed three-quantum-well THz QCLs at various wavelengths featuring different degrees of diagonality, and have analyzed the benefits of diagonal laser transitions for high temperature operation. Between 150 and 250\,K, we find that the main advantage of diagonal structures is the considerably increased inversion; additionally, the reduced gain bandwidth of diagonal transitions is beneficial. These effects can outweigh the reduced oscillator strength and provide advantage in QCL temperature performance.  For designs operating above approximately 3.5 THz, we find that diagonal structures offer advantages at operating temperatures below 200\,K. For lower frequencies the advantages offered by diagonal designs become relevant only at operating temperatures of 200-250\,K or higher. These simulation results provide a basis for a further optimization of the temperature performance of terahertz QCLs.

A. M. and C. J. acknowledge support from the Emmy Noether program of the German Research Foundation (DFG, JI115/1-1).
M. B. acknowledges support from the National Science Foundation (ECCS-0935217).


\begin{thebibliography}{10}%
\makeatletter
\providecommand \@ifxundefined [1]{%
 \ifx #1\undefined \expandafter \@firstoftwo
 \else \expandafter \@secondoftwo
\fi
}%
\providecommand \@ifnum [1]{%
 \ifnum #1\expandafter \@firstoftwo
 \else \expandafter \@secondoftwo
\fi
}%
\providecommand \enquote [1]{``#1''}%
\providecommand \bibnamefont  [1]{#1}%
\providecommand \bibfnamefont [1]{#1}%
\providecommand \citenamefont [1]{#1}%
\providecommand\href[0]{\@sanitize\@href}%
\providecommand\@href[1]{\endgroup\@@startlink{#1}\endgroup\@@href}%
\providecommand\@@href[1]{#1\@@endlink}%
\providecommand \@sanitize [0]{\begingroup\catcode`\&12\catcode`\#12\relax}%
\@ifxundefined \pdfoutput {\@firstoftwo}{%
 \@ifnum{\z@=\pdfoutput}{\@firstoftwo}{\@secondoftwo}%
}{%
 \providecommand\@@startlink[1]{\leavevmode}%
 \providecommand\@@endlink[0]{}%
}{%
 \providecommand\@@startlink[1]{%
  \leavevmode
  \pdfstartlink
   attr{/Border[0 0 1 ]/H/I/C[0 1 1]}%
   user{/Subtype/Link/A<</Type/Action/S/URI/URI(#1)>>}%
  \relax
 }%
 \providecommand\@@endlink[0]{\pdfendlink}%
}%
\providecommand \url  [0]{\begingroup\@sanitize \@url }%
\providecommand \@url [1]{\endgroup\@href {#1}{\urlprefix}}%
\providecommand \urlprefix [0]{URL }%
\providecommand \Eprint[0]{\href }%
\@ifxundefined \urlstyle {%
  \providecommand \doi [1]{doi:\discretionary{}{}{}#1}%
}{%
  \providecommand \doi [0]{doi:\discretionary{}{}{}\begingroup
  \urlstyle{rm}\Url }%
}%
\providecommand \doibase [0]{http://dx.doi.org/}%
\providecommand \Doi[1]{\href{\doibase#1}}%
\providecommand \selectlanguage [0]{\@gobble}%
\providecommand \bibinfo [0]{\@secondoftwo}%
\providecommand \bibfield [0]{\@secondoftwo}%
\providecommand \translation [1]{[#1]}%
\providecommand \BibitemOpen[0]{}%
\providecommand \bibitemStop [0]{}%
\providecommand \bibitemNoStop [0]{.\EOS\space}%
\providecommand \EOS [0]{\spacefactor3000\relax}%
\providecommand \BibitemShut [1]{\csname bibitem#1\endcsname}%
\bibitem{2003ApPhL..82.1015W}%
  \BibitemOpen
  \bibfield{author}{%
  \bibinfo {author} {\bibfnamefont{B.~S.}\ \bibnamefont{{Williams}}}, \bibinfo
  {author} {\bibfnamefont{H.}~\bibnamefont{{Callebaut}}}, \bibinfo {author}
  {\bibfnamefont{S.}~\bibnamefont{{Kumar}}}, \bibinfo {author}
  {\bibfnamefont{Q.}~\bibnamefont{{Hu}}},\ and\ \bibinfo {author}
  {\bibfnamefont{J.~L.}\ \bibnamefont{{Reno}}},\ }%
  \bibfield{journal}{%
  \Doi{10.1063/1.1554479}{\bibinfo {journal} {Appl. Phys. Lett.}}\ }%
  \textbf{\bibinfo {volume} {82}},\ \bibinfo {pages} {1015} (\bibinfo {year}
  {2003})\BibitemShut{NoStop}%
\bibitem{2005OExpr..13.3331W}%
  \BibitemOpen
  \bibfield{author}{%
  \bibinfo {author} {\bibfnamefont{B.~S.}\ \bibnamefont{{Williams}}}, \bibinfo
  {author} {\bibfnamefont{S.}~\bibnamefont{{Kumar}}}, \bibinfo {author}
  {\bibfnamefont{Q.}~\bibnamefont{{Hu}}},\ and\ \bibinfo {author}
  {\bibfnamefont{J.~L.}\ \bibnamefont{{Reno}}},\ }%
  \bibfield{journal}{%
  \bibinfo {journal} {Opt. Express}\ }%
  \textbf{\bibinfo {volume} {13}},\ \bibinfo {pages} {3331} (\bibinfo {year}
  {2005})\BibitemShut{NoStop}%
\bibitem{2007ApPhL..90d1112L}%
  \BibitemOpen
  \bibfield{author}{%
  \bibinfo {author} {\bibfnamefont{H.}~\bibnamefont{{Luo}}}, \bibinfo {author}
  {\bibfnamefont{S.~R.}\ \bibnamefont{{Laframboise}}}, \bibinfo {author}
  {\bibfnamefont{Z.~R.}\ \bibnamefont{{Wasilewski}}}, \bibinfo {author}
  {\bibfnamefont{G.~C.}\ \bibnamefont{{Aers}}}, \bibinfo {author}
  {\bibfnamefont{H.~C.}\ \bibnamefont{{Liu}}},\ and\ \bibinfo {author}
  {\bibfnamefont{J.~C.}\ \bibnamefont{{Cao}}},\ }%
  \bibfield{journal}{%
  \Doi{10.1063/1.2437071}{\bibinfo {journal} {Appl. Phys. Lett.}}\ }%
  \textbf{\bibinfo {volume} {90}},\ \bibinfo {pages} {041112} (\bibinfo {year}
  {2007})\BibitemShut{NoStop}%
\bibitem{2009ApPhL..94m1105K}%
  \BibitemOpen
  \bibfield{author}{%
  \bibinfo {author} {\bibfnamefont{S.}~\bibnamefont{{Kumar}}}, \bibinfo
  {author} {\bibfnamefont{Q.}~\bibnamefont{{Hu}}},\ and\ \bibinfo {author}
  {\bibfnamefont{J.~L.}\ \bibnamefont{{Reno}}},\ }%
  \bibfield{journal}{%
  \Doi{10.1063/1.3114418}{\bibinfo {journal} {Appl. Phys. Lett.}}\ }%
  \textbf{\bibinfo {volume} {94}},\ \bibinfo {pages} {131105} (\bibinfo {year}
  {2009})\BibitemShut{NoStop}%
\bibitem{BElkinCapasso2009}%
  \BibitemOpen
  \bibfield{author}{%
  \bibinfo {author} {\bibfnamefont{M.~A.}\ \bibnamefont{{Belkin}}}, \bibinfo
  {author} {\bibfnamefont{Q.~J.}\ \bibnamefont{{Wang}}}, \bibinfo {author}
  {\bibfnamefont{C.}~\bibnamefont{{Pfl\"ugl}}}, \bibinfo {author}
  {\bibfnamefont{A.}~\bibnamefont{{Belyanin}}}, \bibinfo {author}
  {\bibfnamefont{P.~S.}\ \bibnamefont{{Khanna}}}, \bibinfo {author}
  {\bibfnamefont{A.~G.}\ \bibnamefont{{Davies}}}, \bibinfo {author}
  {\bibfnamefont{E.~H.}\ \bibnamefont{{Linfield}}},\ and\ \bibinfo {author}
  {\bibfnamefont{F.}~\bibnamefont{{Capasso}}},\ }%
  \bibfield{journal}{%
  \Doi{10.1109/JSTQE.2009.2013183}{\bibinfo {journal} {IEEE J. Sel. Top. Quantum Electron.}}\ }%
  \textbf{\bibinfo {volume} {15}},\ \bibinfo {pages} {952} (\bibinfo {year}
  {2009})\BibitemShut{NoStop}%
\bibitem{Rossi_New}%
  \BibitemOpen
  \bibfield{author}{%
  \bibinfo{author} {\bibfnamefont{R.~C.}\ \bibnamefont{{Iotti}}}, and\
  \bibinfo {author} {\bibfnamefont{F.}~\bibnamefont{{Rossi}}},\ }%
  \bibfield{journal}{%
  \bibinfo {journal} {Phys. Rev. Lett.}}\ 
  \textbf{\bibinfo {volume} {87}},\ \bibinfo {pages} {146603-1} (\bibinfo {year}
  {2001})\BibitemShut{NoStop}%
\bibitem{CallebautHu}%
  \BibitemOpen
  \bibfield{author}{%
  \bibinfo {author} {\bibfnamefont{H.}~ \bibnamefont{{Callebaut}}},
  \bibinfo {author} {\bibfnamefont{S.}~\bibnamefont{{Kumar}}},
  \bibinfo {author} {\bibfnamefont{B.~S.}~ \bibnamefont{{Williams}}},
  \bibinfo {author} {\bibfnamefont{Q.}~ \bibnamefont{{Hu}}}, and\
  \bibinfo {author} {\bibfnamefont{J.~L.}~\bibnamefont{{Reno}}},\ }%
  \bibfield{journal}{%
  \bibinfo {journal} {Appl. Phys. Lett.}}\ 
  \textbf{\bibinfo {volume} {83}},\ \bibinfo {pages} {207} (\bibinfo {year}
  {2003})\BibitemShut{NoStop}%
\bibitem{BonnoDessenne}%
  \BibitemOpen
  \bibfield{author}{%
  \bibinfo {author} {\bibfnamefont{O.}~ \bibnamefont{{Bonno}}},
  \bibinfo {author} {\bibfnamefont{J.~L.}~\bibnamefont{{Thobel}}}, and\
  \bibinfo {author} {\bibfnamefont{F.}~\bibnamefont{{Dessenne}}},\ }%
  \bibfield{journal}{%
  \bibinfo {journal} {J. Appl. Phys.}}\ 
  \textbf{\bibinfo {volume} {97}},\ \bibinfo {pages} {043702} (\bibinfo {year}
  {2005})\BibitemShut{NoStop}%
\bibitem{AWacker01}%
  \BibitemOpen
  \bibfield{author}{%
  \bibinfo {author} {\bibfnamefont{R.}~\bibnamefont{{Nelander}}}\ and\
  \bibinfo {author} {\bibfnamefont{A.}~\bibnamefont{{Wacker}}},\ }%
  \bibfield{journal}{%
  {\bibinfo {journal} {Appl. Phys. Lett.}}\ }%
  \textbf{\bibinfo {volume} {92}},\ \bibinfo {pages} {081102} (\bibinfo {year}
  {2008})\BibitemShut{NoStop}%
\bibitem{2009JAP...105l3102J}%
  \BibitemOpen
  \bibfield{author}{%
  \bibinfo {author} {\bibfnamefont{C.}~\bibnamefont{{Jirauschek}}}\ and\
  \bibinfo {author} {\bibfnamefont{P.}~\bibnamefont{{Lugli}}},\ }%
  \bibfield{journal}{%
  \Doi{10.1063/1.3147943}{\bibinfo {journal} {J. Appl. Phys.}}\ }%
  \textbf{\bibinfo {volume} {105}},\ \bibinfo {pages} {123102} (\bibinfo {year}
  {2009})\BibitemShut{NoStop}%
\bibitem{2009IJQE...45..1059J}%
  \BibitemOpen
  \bibfield{author}{%
  \bibinfo {author} {\bibfnamefont{C.}~\bibnamefont{{Jirauschek}}},\ }%
  \bibfield{journal}{%
  \Doi{10.1109/JQE.2009.2020998}{\bibinfo {journal} {IEEE J. Quantum
  Electron.}}\ }%
  \textbf{\bibinfo {volume} {45}},\ \bibinfo {pages} {1059} (\bibinfo {year}
  {2009})\BibitemShut{NoStop}%
\bibitem{bandgp}
  \BibitemOpen
  \bibfield{author}{%
  \bibinfo {author} {\bibfnamefont{D.~E.}~\bibnamefont{{Aspnes}}},
  \bibinfo {author} {\bibfnamefont{S.~M.}~\bibnamefont{{Kelso}}},
  \bibinfo {author} {\bibfnamefont{R.~A.}~\bibnamefont{{Logan}}}, and\
  \bibinfo {author} {\bibfnamefont{R.}~\bibnamefont{{Bhat}}},\ }%
  \bibfield{journal}{%
  \bibinfo {journal} {J. Appl. Phys.}}\ %
  \textbf{\bibinfo {volume} {60}},\ \bibinfo {pages} {754} (\bibinfo {year}
  {1986})\BibitemShut{NoStop}%
\bibitem{Chas10}%
  \BibitemOpen
  \bibinfo {author} {\bibfnamefont{Y.}~\bibnamefont{{Chassagneux}}},\
  \bibinfo {author} {\bibfnamefont{J. R.}~\bibnamefont{{Coudevylle}}},\
  \bibinfo {author} {\bibfnamefont{R.}~\bibnamefont{{Colombelli}}},\
  \bibinfo {author} {\bibfnamefont{M. A.}~\bibnamefont{{Belkin}}},\
  \bibinfo {author} {\bibfnamefont{Q. J.}~\bibnamefont{{Wang}}},\
  \bibinfo {author} {\bibfnamefont{R.}~\bibnamefont{{Capasso}}},\
  \bibinfo {author} {\bibfnamefont{S. P.}~\bibnamefont{{Khanna}}},\
  \bibinfo {author} {\bibfnamefont{E. H.}~\bibnamefont{{Linfield}}}, and\
  \bibinfo {author}  {\bibfnamefont{A. G.}~\bibnamefont{{Davies}}},\ \bibinfo {pages}
  {unpublished}\BibitemShut{NoStop}%
\bibitem{2007NaPho...1..517W}%
  \BibitemOpen
\bibfield{author}{%
    }%
  \bibfield{author}{%
  \bibinfo {author} {\bibfnamefont{B.~S.}\ \bibnamefont{{Williams}}},\ }%
  \bibfield{journal}{%
  \Doi{10.1038/nphoton.2007.166}{\bibinfo {journal} {Nat. Photonics}}\ }%
  \textbf{\bibinfo {volume} {1}},\ \bibinfo {pages} {517} (\bibinfo {year}
  {2007})\BibitemShut{NoStop}%
\end{thebibliography}
\end{document}